\documentclass[10pt,a4paper,twocolumn]{article}
\usepackage[utf8]{inputenc}
\usepackage[english]{babel}
\usepackage{amsmath,amsfonts,amssymb}
\usepackage{graphicx}
\usepackage{hyperref}
\usepackage{url,cite,xcolor}
\usepackage{verbatim}
\usepackage{fancyvrb}  
\usepackage{enumitem}  
\usepackage{cuted}     

\DefineVerbatimEnvironment{jsoncode}{Verbatim}{fontsize=\small,frame=single}

\hypersetup{
    colorlinks=true,
    linkcolor=blue,
    filecolor=magenta,      
    urlcolor=cyan,
    pdftitle={Bomfather: An eBPF-based Kernel-level Monitoring Framework},
    pdfauthor={Naveen Srinivasan, Nathan Naveen, Neil Naveen},
    pdfsubject={Computer Science},
    pdfkeywords={eBPF, Software Supply Chain, SBOM},
}

\title{Bomfather: An eBPF-based Kernel-level Monitoring Framework for Accurate Identification of Unknown, Unused, and Dynamically Loaded Dependencies in Modern Software Supply Chains}
\author{\hyperlink{author:naveen}{Naveen Srinivasan} \and \hyperlink{author:nathan}{Nathan Naveen} \and \hyperlink{author:neil}{Neil Naveen}}
\date{\today}

\begin{document}

\maketitle

\begin{abstract}
Inaccuracies in conventional dependency-tracking methods frequently undermine the security and integrity of modern software supply chains. This paper introduces a kernel-level framework leveraging extended Berkeley Packet Filter (eBPF) \cite{6} to capture software build dependencies transparently in real time. Our approach provides tamper-evident, intrinsic identifiers of build-time dependencies by computing cryptographic hashes of files accessed during compilation and constructing Merkle trees based on the observed file content.

\noindent In contrast to traditional static analysis, this kernel-level methodology accounts for conditional compilation, dead-code, selective library usage, and dynamic dependencies, yielding more precise Software Bills of Materials (SBOMs) and Artifact Dependency Graphs (ADGs). We illustrate how existing SBOMs may omit dynamically loaded or ephemeral dependencies and discuss how kernel-level tracing can mitigate these omissions. The proposed system enhances trustworthiness in software artifacts by offering independently verifiable, kernel-level evidence of build provenance, thereby reducing supply chain risks and facilitating more accurate vulnerability management.
\end{abstract}

\section{Introduction}
Modern software supply chains encompass a broad spectrum of open-source components, proprietary libraries, and third-party services, significantly expanding the potential attack surface for developers, organizations, and end-users. Incidents such as the SolarWinds breach \cite{1} and compromises of npm and PyPI packages \cite{2} illustrate how vulnerabilities within the supply chain can lead to widespread compromise. A fundamental issue in addressing these risks is the challenge of trust, as software consumers require reliable and accurate dependency representations through Software Bills of Materials (SBOMs).

SBOM generation often relies on manifest files or static analysis, which may not capture all dependencies employed at compile time or during execution. Conditional compilation, selective linking, dynamic library loading, and ephemeral container layers can collectively cause critical blind spots. Ken Thompson's seminal essay, ``Reflections on Trusting Trust'' \cite{3}, underscores the inherent difficulty of verifying any software artifact one did not produce personally, highlighting the need for robust, verifiable approaches to build provenance.

Although recent initiatives such as Syft \cite{4} and OmniBOR \cite{5} have strengthened software supply chain security via improved provenance and transparency, these solutions typically rely on user-space or build instrumentation. As such, a gap remains between declared and actual dependencies. To address this gap, the tool proposed by this paper is a kernel-level monitoring framework, implemented via extended Berkeley Packet Filter (eBPF) \cite{6}, which provides direct insight into all file accesses during software builds. By generating content-based Merkle trees \cite{7} from these kernel-level observations, the system produces cryptographic ``fingerprints'' that reflect the exact state of the software during compilation.

\section{Problem Statement}
Contemporary software supply chain security efforts rely on static lists of declared dependencies. While such lists can capture libraries specified in package manifests, they often omit key details, leading to several critical security gaps. 

One significant issue is that vulnerabilities usually impact only specific components within a project rather than the whole codebase. Many projects use only parts of their dependencies, potentially avoiding the vulnerable sections entirely. While traditional tools classify entire packages as vulnerable dependencies, our solution pinpoints which components are currently in use. This detailed visibility enables users to evaluate their actual vulnerability exposure, accurately removing unnecessary remediation efforts for components that do not pose any real risk.

Another significant challenge involves dynamic dependencies introduced through interpreted languages, plugin-based architectures, or runtime library loading mechanisms. These dependencies remain hidden from conventional static analysis, creating potential attack vectors that bypass standard security controls. Organizations often remain unaware of these concealed dependencies until they experience unexpected behavior or security incidents.

Containerized builds present an additional complexity, as ephemeral installations within container layers often go unnoticed when analysis focuses solely on the final manifest. These temporary dependencies can introduce vulnerabilities that persist in the final artifact despite not appearing in formal dependency declarations. This gap is particularly concerning as containerization becomes increasingly prevalent in modern software development environments.

Another issue involves conditional compilation, where build flags and environment-specific code may introduce or exclude significant amounts of logic. Static dependency metadata fails to capture these nuances, creating blind spots in security analysis. This limitation becomes particularly problematic when different deployment environments trigger varying code paths with distinct security implications.

\subsection{Consequences of Incomplete Dependency Tracking}
Incomplete dependency tracking introduces critical vulnerabilities into software security frameworks. When systems overlook active dependencies or misrepresent unused libraries, they simultaneously mask serious vulnerabilities and create unnecessary remediation work. Security teams and developers consequently waste valuable resources auditing and patching components that never appear in compiled code or execute during runtime.

Recent regulatory standards, including CISA guidance \cite{8}, now mandate accurate Software Bills of Materials (SBOMs). The lack of comprehensive visibility into actual software composition undermines effective vulnerability prioritization, preventing organizations from implementing targeted mitigation strategies for their most significant vulnerabilities.

\section{Illustrative Incompleteness of Modern SBOMs}
Although traditional SBOM tools like Syft and cdxgen allow for the recording of declared dependencies, they can fail to capture dynamic or runtime dependencies. For example, a Go application may list only the modules present in go.mod, failing to document dynamically loaded plugins. Similarly, container-based builds might install transient packages for compilation before discarding them, which leaves no record in the final SBOM. This leads to a disconnect between declared and actual dependencies, reducing the utility of the SBOM for security auditing.

\subsection{Case Study: Comprehensive SBOM for Go Compiler Build}
To illustrate the depth and complexity of actual build dependencies in modern software, we conducted a kernel-level observation of the Go 1.21.0 compiler build process using our eBPF-based monitoring framework. The build was performed in a containerized environment with a minimal set of declared dependencies in the Dockerfile: git, gcc, libc6-dev, curl, and a bootstrap Go 1.20 compiler.

The kernel-level SBOM generated from this build revealed a substantially more complex dependency graph than would be captured by manifest-based approaches. Our monitoring identified:

\begin{itemize}
    \item 185,184 total file access events during the compilation process
    \item 10,032 Go source files accessed across various stages of compilation
    \item 26 shared object (.so) libraries dynamically loaded during the build
    \item 979 assembly (.s) files processed for architecture-specific implementations
\end{itemize}

The SBOM captured the complete provenance of the build, including kernel-level evidence of each file access with corresponding cryptographic hashes:

\begin{strip}
\begin{center}
\begin{jsoncode}
{
  "type": "file",
  "name": "bomfather:/go-source/src/internal/platform/supported.go",
  "hashes": [
    {
      "alg": "SHA-256",
      "content": 
      "fe8b88d8b412ba7119e6f37a00415faec9923b7f379561330dadfb4758b43c4b"
    }
  ],
  "properties": [
    {
      "name": "bomfather:pid",
      "value": "93844"
    }
  ]
}
\end{jsoncode}
\end{center}
\end{strip}

Beyond individual files, our approach documented the precise commands executed during the build process with their complete environment variables:

\begin{strip}
\begin{center}
\begin{jsoncode}
"properties": [
  {
    "name": "bomfather:command:pid=81530",
    "value": "runc:[2:INIT] ./make.bash          
        Env: PATH=/usr/local/sbin:/usr/local/bin:/usr/sbin:/usr/bin:
        /sbin:/bin:/usr/local/go/bin:/go/bin, 
        HOSTNAME=37ef788854ed, GOPATH=/go, HOME=/root, , , , , , "
  },
  {
    "name": "bomfather:command:pid=82207",
    "value": "dist /usr/local/go/bin/go install 
        -tags=math_big_pure_go compiler_bootstrap purego 
        bootstrap/cmd/...
        Env: HOSTNAME=37ef788854ed, 
        GOROOT_BOOTSTRAP=/usr/local/go, 
        HOME=/root, 
        DIST_UNMODIFIED_PATH=/usr/local/sbin:/usr/local/bin:
        /usr/sbin:/usr/bin:/sbin:/bin:/usr/local/go/bin:/go/bin, 
        GOROOT=/usr/local/go, SHLVL=1, 
        PATH=/go-source/bin:/usr/local/sbin:/usr/local/bin:/usr/sbin:
        /usr/bin:/sbin:/bin:/usr/local/go/bin:/go/bin, 
        GOPATH=/go-source/pkg/bootstrap, _=./cmd/dist/dist, TERM=dumb"
  }
]
\end{jsoncode}
\end{center}
\end{strip}

Particularly significant was the capture of system-level shared objects that would typically be invisible in traditional SBOMs:

\begin{strip}
\begin{center}
\begin{jsoncode}
{
  "type": "file",
  "name": "bomfather:/usr/lib/gcc/aarch64-linux-gnu/11/libgcc_s.so",
  "hashes": [
    {
      "alg": "SHA-256",
      "content": 
      "69a56a9993b7729b29b274e65016031c81f2397f176ed5ad44d59bd50425e0bd"
    }
  ],
  "properties": [
    {
      "name": "bomfather:pid",
      "value": "90421"
    }
  ]
}
\end{jsoncode}
\end{center}
\end{strip}

The SBOM also captured architecture-specific assembly files that are conditionally included based on build targets:

\begin{strip}
\begin{center}
\begin{jsoncode}
{
  "type": "file",
  "name": "bomfather:/go-source/src/runtime/rt0_openbsd_arm.s",
  "hashes": [
    {
      "alg": "SHA-256",
      "content": 
      "b89ee998ebe14d1f69ede3dfd3e698c5844b6379b81d206aa5d76ca0f20644f3"
    }
  ],
  "properties": [
    {
      "name": "bomfather:pid",
      "value": "93814"
    }
  ]
}
\end{jsoncode}
\end{center}
\end{strip}

This comprehensive SBOM is anchored by a Merkle root hash that serves as a cryptographic fingerprint of the entire build process:

\begin{strip}
\begin{center}
\begin{jsoncode}
{
  "name": "bomfather:merkle_root",
  "value": 
  "b0a1471275b547823ad4ccf778c7eb2e471c2c9960adfb678eebc11cab5115e4"
}
\end{jsoncode}
\end{center}
\end{strip}

This case study demonstrates that the actual dependencies in a software build can be orders of magnitude more complex than those captured in traditional SBOMs. Without kernel-level visibility, security teams would miss critical dependencies that could harbor vulnerabilities, particularly those introduced through dynamically loaded libraries, conditionally compiled assembly code, and transient build artifacts.

\section{Our Approach}

\subsection{Kernel-Level Monitoring with eBPF}
Our system employs extended Berkeley Packet Filter (eBPF) \cite{6} to observe file operations at the kernel level. This method transparently collects real-time data on every file accessed during compilation by attaching probes to system calls such as open() and read(). Each log entry includes the file path, process ID, and a cryptographic hash (i.e., SHA-256) \cite{9} of the file contents. This capture is independent of user-space or build-tool instrumentation, reducing the risk of missed dependencies.

\subsection{Real-Time Dependency Tracking and pURL Generation}
During the build process, we hash each captured file and generate a software identifier that conforms to the Package URL (pURL) specification \cite{10}. The pURL ensures a standardized format for referencing each component, facilitating interoperability with vulnerability scanners and compliance tools. The direct kernel-level capture also records critical metadata, such as environment variables and command line arguments, enabling a more accurate depiction of the build environment.

The pURLs are generated directly from the files monitored by eBPF, allowing end users to verify these pURLs. This verifiable link connects intrinsic file-level identifiers with extrinsic package-level identifiers.

\subsection{Content-Based Merkle Tree Generation}
A central element of this framework is the construction of content-based Merkle trees \cite{7} from all file hashes. Merkle trees enable:

\begin{itemize}
    \item \textbf{Tamper-Evident Verification:} Any alteration to a leaf node (file hash) cascades upward, invalidating the root hash.
    \item \textbf{Intrinsic Build Identification:} The Merkle root is a cryptographic fingerprint of the entire build.
    \item \textbf{Efficient Comparison:} Comparing Merkle roots across builds quickly detects changes, with partial proofs pinpointing modifications at the file level.
\end{itemize}

\section{Addressing the Trust Problem in SBOMs}
Ken Thompson's ``Reflections on Trusting Trust'' \cite{3} highlights the fundamental difficulty of verifying software artifacts one did not create. Malicious actors or misconfigurations can manipulate SBOMs by hiding specific files or entire components. Our approach allows any system to verify the records of the build process and observe which files were accessed via a Merkle tree of content hashes. As the foundation of Zero Trust Architecture, this verification requirement ensures no component receives implicit trust. Consequently, discrepancies between the SBOM and the actual build-time logs reveal potential omissions or adulterations, offering a robust mechanism to establish trust.

\section{Key Benefits}
Our approach advances software security through precise dependency identification. We overcome limitations inherent in static manifests by capturing both compile-time and runtime dependencies while recording actual compilation outcomes—including ephemeral build steps in containerized environments. The system produces Merkle-based cryptographic signatures that simplify unauthorized modification detection and ensure tamper-evident artifacts.

This methodology significantly enhances transparency and regulatory compliance by mapping each SBOM entry to kernel-level evidence. Security teams benefit from focusing exclusively on components actively used in production, eliminating wasted effort on irrelevant dependencies. Additionally, our framework embeds reproducibility anchors such as environment variables and command line arguments, ensuring that identical build inputs consistently produce identical outputs. This is a cornerstone of reliable software verification.

\section{Integration with Existing Development Processes}

\subsection{Compiler and Build-System Compatibility}
The described methodology is compatible with widely used compilers (e.g., GCC \cite{11}, Clang \cite{12}, Go \cite{13}) and builds systems, as it merely observes kernel-level calls without modifying compiler source code. For interpreted languages (e.g., Python \cite{14}, JavaScript \cite{15}), kernel-level tracing of script execution and module loading captures runtime dependencies.

\subsection{CI/CD Pipelines and Containerized Environments}
The approach naturally extends to continuous integration and deployment pipelines. Container-based workflows can attach eBPF probes in ephemeral environments, producing precise SBOMs for each build iteration. This integration promotes transparency and traceability, aligning automated builds with rigorous supply chain security practices.

\subsection{Post-Build Packaging and Distribution}
The resulting cryptographic data (e.g., Merkle root hashes) can be embedded directly within software artifacts or accompanying metadata. The Merkle root allows downstream consumers to verify artifact integrity by comparing embedded hashes against a reference ledger or transparency log.

\section{Conclusion}
This paper presents a kernel-level build validation framework leveraging extended Berkeley Packet Filter (eBPF) and content-based Merkle trees. By transparently monitoring file operations at compile time and linking them to cryptographic identifiers, our approach addresses crucial limitations in conventional SBOM generation—namely, the omission of dynamically loaded libraries, conditionally compiled code and other ephemeral dependencies.

The resulting tamper-evident SBOMs and Artifact Dependency Graphs enable independent verification of declared versus actual build-time dependencies, mitigating supply chain risks and offering a robust mechanism to ``trust the trust,'' as articulated by Ken Thompson \cite{3}. We conclude that kernel-level observability and well-structured cryptographic evidence substantially enhance modern software supply chain security practices.

\section*{About the authors}
\hypertarget{author:naveen}{\textbf{Naveen Srinivasan}}\\
\url{https://github.com/naveensrinivasan}

\hypertarget{author:nathan}{\textbf{Nathan Naveen}}\\
\url{https://github.com/nathannaveen}

\hypertarget{author:neil}{\textbf{Neil Naveen}}\\
\url{https://github.com/neilnaveen}

\end{document}